
\normalspace
\overfullrule=0pt

\def\ket #1{\left\vert#1\right\rangle}
\def\aver #1{\left\langle#1\right\rangle}

\FRONTPAGE
\line{\hfill BROWN-HET-936}
\line{\hfill February 1994}
\bigskip
\title{{\bf
OFF-SHELL AMPLITUDES IN TWO DIMENSIONAL OPEN STRING FIELD
THEORY
     \foot{Work Supported in part by
the Department of Energy under
contract DE-FG02-91ER40688 -- Task A.}
     }}
\bigskip
\centerline{Branko Uro\v sevi\'c}
\centerline{\it Physics Department, Brown University, Providence, RI 02912,
USA}
\bigskip
\abstract
{In this note we present an explicit procedure for the regularization of tree
level amplitudes involving discrete states, using open string field theory. We
show that there is a natural correspondence between the discrete states and
off--shell states, later acting as a regularized version of the former. A
general off-shell state corresponds to several physical states. In order to
obtain the well--defined $S$ matrix elements one has to choose representatives
close but not equal to the desired values of external momenta. The procedure
renders finite all $4$-point amplitudes with an even number of (naively)
divergent channels even after the regularization is removed. The rest of the
amplitudes can be defined by means of such regularization.
}

\vfill
\endpage

\chapter{{\bf Motivation and Introduction}}
During the last couple of years $d \leq 1$ string models received considerable
attention, Ref. 1. The main reason for that is that they provide examples of
consistent and exactly soluble string models. Recently, it has been pointed out
that one can construct a non--trivial, consistent theory of strings with
Dirichlet boundary conditions from a collection of $2d$ topological gravity
models, Ref. 2. Dirichlet strings, on the other hand, have been argued to be a
possible candidate for an effective QCD string theory, Ref. 3. In a different
approach toward QCD strings, Gross and Taylor showed the equivalence between
$2d$ QCD and an effective $2d$ string theory without folds (discrete states),
Ref. 4. It is desirable to much better understand various $2d$ models and
relations between them.

In this note we focus on open $2d$ strings. It should be remarked that there is
no matrix model available in that case. The only results known until recently
were
due to Bershadsky and Kutasov who calculated the bulk amplitudes for
tachyon--tachyon scattering in the continuum (Liouville) approach, Ref. 5.
Although simple (the only {\it field} in two dimensional string theory is a
massless ``tachyon"), this is not a trivial theory. It possesses a large
spectrum--generated $W_{\infty}$ symmetry. The generators of the symmetry,
discrete states (DS), are defined only for some particular values of momenta.
In the framework of string field theory , DS appear as BRST cohomology classes,
as well as poles of the tree level $S$ matrix, Ref. 6, 7. In that respect, they
are remnants of higher (excited) string states in $2d$. However, when treated
as external (asymptotic) states, presence of a sufficient number of DS leads to
tree level divergences.

 A classification of $4$--point amplitudes involving DS was presented in Ref. 7
(see Sec. 2 for more details). A $4$-point amplitude diverges if at least one
of the kinematic invariants $s$, $t$ or $u$ is non--positive integer. If $s$ is
such an integer, for example, then the $s$ channel amplitude $\sum_{n \ge 0} \,
{A_n(t)
 \over {s + n}}$ is  ill--defined because the denominator blows up for the
level $n = -s$. If $s > 0$, on the other hand, the amplitude vanishes.

The aim of this note is to treat the divergences more carefully using field
theory. We proceed in a simple way. First we generalize the concept of DS and
allow them to depart from the mass shell in order to render the tree amplitudes
finite. Such a procedure gives a well--defined result. One can think of such
off--shell ``discrete states"  as just a regularization scheme. Then we show
that, upon such regularization, some divergences cancel each other. Such
amplitudes are well--defined even after the regularization is removed. Our
considerations may be generalized to $N$ point amplitudes.

\chapter{{\bf Green's Functions and $S$ Matrix}}

The starting point in our discussion is Witten's open string field theory
(OSFT) in $2d$. As in any other gauge theory, to derive Green's functions in
$2d$ OSFT one needs first to fix the gauge. We choose the so called Siegel's
gauge, $b_0 \, A = \, 0$ (see Ref. 6). Introducing the external sources $J_s$,
one for each coefficient field, one determines the generating functional
$Z(J_s)$ from which
it is easy to deduce (off--shell) Green's functions:

$$
Z(J_s) \, \propto \, \int \, {1 \over n!} \, G_n \, (p_1, \cdots \, , p_n) \,
J_n \, \cdots \, J_1  \, ,
\eqno\eq
$$

\noindent
 where $G_n$ contain an overall factor $\delta^{(2)} \, ( \, \sum_{i=1}^n \,
p_i \, + \, Q\, )$, and $Q^{\mu} = \, (0, 2 \sqrt 2)$ is the background charge
of the matter--Liouville system. We need, however, to specify what is meant by
off--shell states and how they relate to the physical states of the theory. Let
us recall, first,  that $H^{(1)}$ DS can be obtained by applying the raising or
lowering $SU(2)$ charges $H_{\pm} = {1 \over {2 \pi i}}
 \, \oint \, e^{\pm i \sqrt 2 \, x}$  to discrete tachyons $V_{s,\, s}^{\pm} =
\, c \, e^{i \sqrt 2\, s \, x}\, e^{(- \sqrt 2\, \pm \, \sqrt 2 \, s )\,
\varphi}$:

$$  \eqalign
{
$W_{s,\, n}^{\pm}\,  &\propto \, (H_{-})^{s-n} \, V_{s, \, s}^{\pm} \propto
\cr
 &\propto \, \epsilon_{i_1 \cdots \, i_k} \, \alpha^{i_1}_{-j_1}  \, \cdots \,
\alpha^{i_k}_{-j_k} \, \ket{\sqrt 2 n, -\sqrt 2 \pm \sqrt 2 s} \,  \cr ,
}
\eqno\eq
$$

\noindent
where $\epsilon_{i_1 \cdots \, i_k} \propto \, \sum_{j=0}^k \, c_j
\delta_{i_1,1} \, \cdots \delta_{i_j.1}$ is the polarization tensor, and $\sum
k \, j_k$ is the level of the state $W_{s,\, n}^{\pm}$. In this representation
only the matter field contributes to excitations. The set of physical states
can be partitioned with respect to the following equivalence relation: two
states are considered equivalent if they have the same polarization structure
and differ only by the values of momenta. For example, discrete and generic
tachyons belong to one class, vector particles to another and so on. Denote the
quotient space by {\it P}. Let us introduce, now, another set {\it O} which
consists of states with the same polarization stucture as physical states but
with unphysical values of momenta. We refer to the elements of {\it O} as {\it
off--shell} states. It is clear that there exists a natural bijective map {\it
P} $\mapsto$ {\it O}. We can visualize an element of {\it O} as an open,
connected subset of ${\cal

R}^2$, with discrete (physical) states corresponding to points cut out from the
surface (Fig. 1). Note that, in general, more than one physical state maps to
the same off-shell ``surface".

 Tree level fur point Green's functions are given by the following expression:

$$
G \,  \propto \, \prod_{i=1}^4 \, D_i \, \sum_n \, V_{2 n 1} \, D_n \, V_{4 n
3} \, .
\eqno\eq
$$

\noindent
Here, $D_i$ represent external propagators (``legs"), sum is over a complete
set of intermediate states and each of the states $i$ belongs to {\it O}. To
recover (on--shell) $S$ matrix elements from the expression for Green's
function $(2.3)$ one needs to: a) Cut the external legs; b) For each  external
leg choose an on--shell representative. From the comment above it is clear that
one off-shell Green's function corresponds to a variety of $S$ matrix elements.

Let us consider, for example, tachyon Green's function. Contribution to the $s$
channel reads:

$$
\aver{\phi (p_1) \, \cdots \, \phi(p_4)}  \, \propto \, g^2 \, \,
\prod_{i=1}^4 {1 \over \mu_i^2} \, \,  ({4 \over 3 \sqrt 3})^{\sum_{i=1}^4 \,
\mu_i^2} \,
{A_n \, (t,u) \over s + n} \, ,
\eqno\eq
$$

\noindent
where $\mu_i^2 = \, {1 \over 2} \, (p_i + {Q \over 2})^2}$ is $i$-th tachyon
external leg and the first couple of residues are given by:

$$    \eqalign
{
&A_0 = \, 1, \cr
&A_1 = \, {1 \over 2} \, (t - u) \, , \cr
&A_2 =  \, - {1 \over 8} + \, {1 \over 8} (t -u)^2 \, - {5 \over 32} \,
\sum_{i=1}^4 \, \mu_i^2 \cr
}
\eqno\eq
$$

The total Green's function is: $G^{(tot)}  = \, G^{(s)} + G^{(t)}+ G^{(u)}$. To
obtain $S$ matrix elements, after we calculate the residues in $\mu^2$ we need
to pick the values of tachyon (on--shell) momenta we are interesting in. The
same procedure can be repeated for the  non--tachyonic states. In this way one
can reproduce the results of Ref. 7. Let us briefly summarize them. If we
denote by $T$ a generic tachyon and by $D$ an arbitrary DS including discrete
tachyons, then: a) Classes $A_{TTTT}$ and $A_{TDDD}$ are empty; b) Class
$A_{TTTD}$ is well-defined and the amplitudes have an infinite number of
intermediate physical states;  c) Class $A_{TTDD}$ has at least one degenerate
channel and it can be subdivided into three subclasses: $A_{TTDD}^{+}$ is well
defined and has a finite number of intermediate states, $A_{TTDD}^{-}$ diverges
and has an infinite number of intermediate states, while $A_{TTDD}^{deg}$ is
just a divergent number; Finally, $A_{DDDD}$ always diverges. Let us stress,
however, that bein

g off-shell, even not far from the mass shell, makes all amplitudes
well--defined. Moreover, as we shall see in the next section, if an amplitude
is apparently divergent in two channels, these divergences can be made to
cancel each other. Such amplitudes are well--defined even after the
regularization is removed.

\chapter{\bf Off--Shell Amplitudes and Regularization}

In the previous section we have defined off--shell states and discussed their
relationship with physical states. We have seen that several physical states
correspond to the same off--shell state. Green's functions $(2.3)$ are
well--defined off-shell. In fact, none of the kinematic invariants $s$, $t$ or
$u$ degenerate because, apart from the overall momentum conservation, there is
no other kinematic constraints. Cutting the external legs or, in other words,
calculating the residues in $\mu_i^2$, gives rise to the constraint $s + t + u
= \, \sum_{i=1}^4 \, \mu_i^2 \, \leq 1$, which is still harmless. The only
potentially harmful step in the procedure of $S$ matrix calculation is the last
step, namely, choosing the representatives of the external states. This is what
we shall discuss next.

Let us consider a specific example, namely, the amplitude belonging to the
class $A_{TTDD}^{(deg)}$:

$$
\aver{W_{{1 \over 2}, {1 \over 2}}^{+} \, W_{k_2}^{-} \, W_{k_3}^{-} \, W_{{1
\over 2}, {1 \over 2}}^{+}} = \, \int_0^1 \, {dx \over x}
\eqno\eq
$$

\noindent
Notice that $(3.1)$ is non--dynamical (momentum--independent), although
divergent. One would expect that it does not contribute to the {\it properly}
defined scattering amplitude. Let us show that this is, indeed, the case. We
start from the off--shell tachyon Green function $(2.4)$. Matter and Liouville
momenta are, at this point, arbitrary (subject, of course, to the conservation
law). Let us, now, pick the values of momenta close to the physical values
indicated in $(3.1)$, namely:

$$  \eqalign
{
&p_1^{\mu} = \, ({1 \over \sqrt 2} + \epsilon_1^m, \, - {1 \over \sqrt 2} +
\epsilon_1^l) \, , \cr
&p_2^{\mu} = (k_2 + \epsilon_2^m, \, - \sqrt 2\, - k_2 \, + \epsilon_2^m) \, ,
\cr
&p_3^{\mu} = (k_3 + \epsilon_3^m, \, - \sqrt 2\, - k_3 \, + \epsilon_3^m) \, ,
\cr
&p_4^{\mu} = \, ({1 \over \sqrt 2} + \epsilon_4^m, \, - {1 \over \sqrt 2} +
\epsilon_4^l) \, , \cr
}
\eqno\eq
$$

\noindent
{}From the momentum conservation, one has a constraint on the allowed values of
$\epsilon_i$'s:

$$
\sum_{i=1}^4 \, \epsilon_i^m \, = \, - \sum_{i=1}^4 \, \epsilon_i^l \, .
\eqno\eq
$$

\noindent
which is identically satisfied if we choose: $\epsilon_i^m = \epsilon = -
\epsilon_i^l$. Selecting a particular regularized $S$ matrix element is
equivalent to choosing sufficiently small, non-intersecting neighborhoods
around those value of momenta (Fig. 2). Such neighborhoods always exist since
each element of {\it O} is a Hausdorff topological space. The neighborhoods are
off--shell representatives (regularizations) of the on--shell state. It follows
that the choice made in $(3.2)$ gives a particular contribution to the $S$
matrix only if we restrict ourselves to the terms small in $\epsilon$. The
limit $\epsilon \rightarrow 0$, if exists, gives then the value of the
particular on--shell amplitude. For the larger values of $\epsilon$, on the
other hand,  neighborhoods start overlapping with each other. That situation
corresponds to a {\it generic} off--shell amplitude. Below, we discard all
quantities of the order $o ( \, \epsilon \, )$.

Kinematic invariants, then, read:

$$   \eqalign
{
&s = {1 \over 2} \, (p_1 + p_2 + {Q \over 2} )^2 \, \simeq \, 4 \, \epsilon \,
({1 \over \sqrt 2} + k_2) \, , \cr
&t = {1 \over 2} \, (p_1 + p_3 + {Q \over 2} )^2 \, \simeq \, - 4 \, \epsilon
\, ({1 \over \sqrt 2} + k_2) \, \simeq \, - s , \cr
&u = {1 \over 2} \, (p_1 + p_4 + {Q \over 2} )^2 \, \simeq \, 1 + 2 \sqrt 2  \,
\epsilon . \cr
}
\eqno\eq
$$

\noindent
We see that in the limit $\epsilon \rightarrow 0$, $s$, $t$ $\rightarrow 0$,
while $u \rightarrow 1$. This means that, had we altogether neglected
$\epsilon$ we would have two divergent channels, divergences arising from the
$n=0$ (tachyonic) intermediate state. Let us keep, for now, {\it finite}
$\epsilon$ and consider the potentially divergent piece of the {\it total}
amplitude more carefully. We have:

$$
A^{(tot)} \simeq \, {{({27 \over 16})^s \, - \, ({27 \over 16})^{-s} } \over
s}.
\eqno\eq
$$

\noindent
We see that the residue in $s$ vanishes, so that the amplitude has, in fact no
singularities. Now we can take the limit $\epsilon \rightarrow 0$. Evidently,
$A^{(deg)} = 0$. This result can be generalized to any $4$-point amplitude
which (naively) diverges in {\it two} different channels. The crucial point
here is that two divergent contributions (from two different channels) can
always be made to cancel each other. In the case of an odd number of divergent
channels this simple argument does not work. This is the case, for example, for
$A_{TTDD}^{(-)}$. There we only have one divergent channel, so that the residue
of the divergent piece is not zero and the cancellation does not occur.
Similarly,  some of the amplitudes $A_{DDDD}$ have three divergent channels.
(For example, for
$\aver{ W_{{3 \over 2},-{1 \over 2}}^{+} \, W_{{5 \over 2},{1 \over 2}}^{+} \,
W_{1, 0}^{-} \,
W_{1, 0}^{-}}$ one has $s = - 9$ and $t = u = 0$). It is easy to see that
divergences of two of the channels may cancel each other but there will be
always one left. It is not a disaster, however. In that case we could keep
$\epsilon_i$ finite, and have well--behaved expressions which we can {\it
define} to be the answer we are looking for.

So, to summarize, departing from the mass shell gives a natural way to
regularize the amplitudes involving discrete states since the states are not,
basically, discrete any more. In the case of an even number of divergent
channels divergences cancel each other and the amplitude is well defined even
after the regularization is removed. For the amplitudes with an odd number of
divergent channels one can {\it redefine} them using the procedure above.

It is easy to see that the similar situation arises in general $N$-point case.
As an illustration, consider $5$-point Green's function. It is given by:

$$
G \propto \, g^3 \, \prod_{i=1}^5 \, D_i \, V_{2 K 1} \, D_K \, V_{K  3 L} \,
V_{5 L 4} \, ,
\eqno\eq
$$

\noindent
plus nonequivalent permutations. Upon chopping the external legs we are left
with
two intermediate propagators. Each of them, potentially, could give rise to a
divergence. It is going to be the case always when a sufficient number of DS is
involved. Departing from the mass shell makes amplitudes well--defined and the
procedure similar to the one developed for the four point functions may be
applied.

\chapter{\bf Final remarks}

In this note we suggested a simple way to deal with the divergences of the tree
level amplitudes involving discrete states. Namely, we suggested to use the
off-shell formulation of the $2d$ OSFT. We have seen that there is a natural
correspondence between the discrete states and off--shell states, later acting
as a regularized version of the former. To obtain well--defined $S$ matrix
elements one, in general, needs to choose off--shell states close but not equal
to the desired values of external momenta (somewhat analogous to the way in
which in Minkowski space one replaces $m^2 \mapsto m^2 - \, i \epsilon$). We
have shown that the procedure renders finite all $4$-point amplitudes with an
even number of (naively) divergent channels even after the regularization is
removed. The rest of the amplitudes can be defined by means of such
regularization.

Discrete states are consequence of a gauge theory in two dimensions.
If the loop corrections are taken into account DS might be  ``broadened". In
other words, instead of a monochromatic wave one would have a wave packet.
If $2d$ string theory has anything to do with some two dimensional condensed
matter system, which is not totally unreasonable assumption (Ref. 8), this kind
of broadening of the ``resonance" spectrum would be expected. It is interesting
to speculate that these  ``dressed" states might be related to the off--shell
states.

Recently, $2d$ closed string field theory has been formulated, Ref. 9. Although
the formulation of the theory is much more complicated that in the open string
case, there seems to be no principal difficulties in treating the divergences
arising there in exactly the same way.

Finally, note that we have considered here discrete states as asymptotic string
states. This is not the only possibility. As an alternative, one may consider
scattering of tachyons in various discrete state backgrounds (similar project
was carried out for the black hole background in  Ref. 10). Another compelling
approach is to consider a field theory of a tachyon coupled to the complete set
of two dimensional topological gravity states. There are indications that such
an approach might shed some light on relation between string field theory and
$W_{\infty}$ (author would like to thank Z. Qui for discussions on this
matter). These and other interesting problems shall be discussed elsewhere.

\noindent
{\bf Acknowledgments.} I am grateful to A. Jevicki for his
guidance and support throughout the work on this project
 and to Z. Qui and M. Li for valuable discussions.

\vfill
\endpage

\centerline{\it REFERENCES}
\bigskip

\point
A. Jevicki, `Developments in $2d$ String Theory', Lectures given at the
 Spring School on String Theory, Trieste, April 1993, Brown HET--918.

\point
Z. Qui, {\it Phys. Lett.} {\bf B306} (1993) 261.

\point
M.B. Green and J.A. Shapiro, {\it Phys. Lett.} {\bf B64} (1976) 454;
M. Li, `Dirichlet String Theory and Singular Random Surfaces', BROWN-HET-923.

\point
D. Gross and W. Taylor, {\it Nucl.Phys.} {\bf B403} (1993) 395.

\point
M. Bershadsky and D. Kutasov, {\it Phys. Lett.} {\bf B274} (1992) 331;
PUPT-1315, HUPT-92/A016.

\point
B. Uro\v sevi\'c, {\it Phys. Rev. D15} {\bf 47} (1993) 5460.

\point
B. Uro\v sevi\'c, {\it Phys. Rev. D15} {\bf 48} (1993) 5827.

\point
B. Marston, Private Communications.

\point
M. Kaku, `Symmetries and String Field Theory in $D=2$', CCNY-HEP-93-5;
M. Kaku, `Subcritical Closed String Field Theory in $d , 26$', CCNY-HEP-93-6.

\point
A. Jevicki, T. Yoneya, `A Deformed Matrix Model and the Black Hole Background
in Two--Dimensional String Theory', NSF-ITP-93-67, BROWN-HEP-904.

\endpage

\end